\documentclass[fleqn,twoside]{article}
\usepackage{espcrc2}
\usepackage{epsfig}

\usepackage{graphicx}
\usepackage[figuresright]{rotating}

\newcommand{\bi}{\begin{itemize}}
\newcommand{\ei}{\end{itemize}}

\newcommand{\AmS}{{\protect\the\textfont2
  A\kern-.1667em\lower.5ex\hbox{M}\kern-.125emS}}

\title{\LARGE{CompHEP 4.4 ~-~ Automatic Computations
from Lagrangians to Events}}

\author{E.~Boos, V.~Bunichev, M.~Dubinin, L.~Dudko, V.~Ilyin, 
A.~Kryukov, V.~Edneral, V.~Savrin, A.~Semenov, A.~Sherstnev\\
(the CompHEP collaboration)
 \address{Skobeltsyn Institute of Nuclear Physics, Moscow
State University, 119992 Moscow, Russia}}
       
\begin{document}

\begin{abstract}
We present a new version of the CompHEP program (version 4.4). We describe
shortly new issues implemented in this version, namely,
simplification of quark flavor
combinatorics for the evaluation of hadronic processes, Les Houches Accord
based CompHEP-PYTHIA interface, processing the color configurations of
events,
implementation of MSSM, symbolical and numerical batch modes, etc.
We discuss how the CompHEP program is used for preparing event generators
for various physical processes.
We mention a few  concrete physics examples for CompHEP based generators
prepared for the LHC and Tevatron.

\vspace{1pc}
\end{abstract}

\maketitle

\section{INTRODUCTION}

The increase of collider energies requires efficient 
and precise calculation
of processes with multi-particle final states. The NLO, NNLO
or resummation corrections to the leading order results
should be incorporated when available.

At LEP1 the most interesting physical channels were 2-fermion 
production processes at the $Z$ pole, while  
at LEP2 mainly 4-fermion processes with some photons or gluons
radiated off have been considered. At the Tevatron, LHC and
Next Linear Colliders a majority of interesting channels include
5,6 and 8 or even more fermions in the final states 
with additional photons or gluons.
For example, the top quark pair production leads to 6 fermions
in the final state,
single top production in various modes gives 4, 5 and 6 fermions,
the channel $t\bar{t}H$ produces 8 fermions in the final state, etc. 
Calculations of event
characteristics with multi-particle final states are needed to be done
at the complete matrix element level in order to include correctly
the spin correlations, nontrivial kinematics and in some cases finite 
width effects in a gauge invariant manner.

In practical computations, especially for hadron colliders, there are
many contributing Feynman diagrams even at the tree level and a large 
number of subprocesses leading to the same multi-particle final state. 
To perform such computations a number of automatic programs have been created~-~ 
CompHEP \cite{Pukhov:1999gg}, GRACE \cite{Fujimoto:2002sj}, 
MadGraph \cite{Stelzer:1994ta},
 AlpGen \cite{Mangano:2002ea}, O' Mega \cite{Moretti:2001zz} 
 with WHIZARD \cite{Kilian:2002cg},
 Amegic \cite{Krauss:2001iv},  HELAC-PHEGAS \cite{Kanaki:2000ms} and 
other.

However, in order to get reliable predictions and reduce
a dependence on scales involved, like QCD 
normalization and factorization scales,
one should compute and simulate not only processes with many final particles
but also include various corrections and perform resummations
of large logs. Recently a few calculations of NLO corrections for 
$2\rightarrow 3$ processes have been performed 
\cite{Campbell:2003hd,Beenakker:2002nc,Dawson:2003zu,Belanger:2003nm}. 

Precise calculations including the NLO corrections to the rate and
distributions are however not sufficient for experimental studies
insofar as one should also provide a corresponding NLO event generators.
Since the real final states are not partons but hadrons (jets) 
these generators should be matched in some way to the parton shower 
generators, such as PYTHIA \cite{Sjostrand:2003wg} and HERWIG 
\cite{Corcella:2000bw}, which simulate a realistic picture 
of hadronization. In fact the problem of matching of all the needed 
ingredients, multi-particle matrix element calculations with all spin 
correlations, NLO corrections to, at least, basic production processes, 
hadronization taking into account initial and final state radiation
is highly nontrivial. As a result one likes to have
an event generator which includes all these parts without "double
counting" of contributions and with smooth distributions in the whole 
range of the phase space. Although this problem is far from a final 
solution, recently interesting and promising approaches for different
parts of the above mentioned matching have been proposed 
\cite{Boos:2001cv,Catani:2001cc,Frixione:2002ik,Frixione:2004wy,Mrenna:2003if,Mangano:2003ps}. 

In this paper we discuss mainly a new aspects of the CompHEP latest 
version 4.4 and present some recent CompHEP applications
to physics studies at colliders. We discuss shortly the SingleTop
generator, created with the help of CompHEP to simulate the electroweak 
top quark production at the Tevatron and LHC, where proper matching of
different processes with main contributions to different phase space 
regions is performed in the final event flow.

CompHEP is based on quantum theory of gauge fields and includes
the Standard Model Lagrangian in the unitary and t'Hooft-Feynman
gauges as well as several other MSSM based models. CompHEP is able to
compute basically the LO cross sections
and distributions but with many particles (up to 4-6) in the final state
taking into account, if necessary, all the QCD and EW diagrams, masses of
fermions and bosons and widths of unstable particles. 
Processes computed by means of CompHEP
could be used as a "new" external processes for generators
like PYTHIA.

\begin{figure}[hbtp]
\includegraphics[width=70mm,height=50mm]{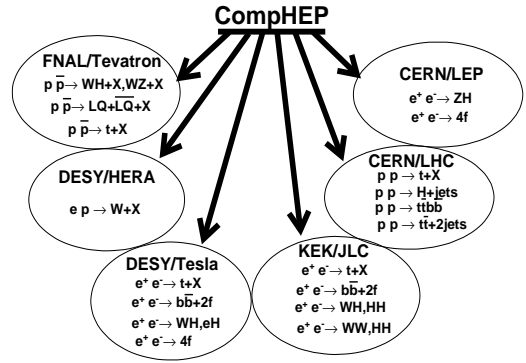}
\vspace{-8mm}
\caption{Very incomplete list of processes computed by means of 
CompHEP in the past}
\label{fig:comp_us}
\end{figure}

The CompHEP project started in 1989 at SINP MSU. During the
90's the package was developed, and now it is a powerful tool for automatic
computations of collision processes. 
The CompHEP program has been used in the past for many studies 
(see \cite{Baikov:1997zr} and the citation to CompHEP \cite{Pukhov:1999gg} 
for more complete information)
as shown schematically in the Fig.\ref{fig:comp_us}. 

%
\section{COMPHEP 4.4}
In the latest CompHEP version 4.4 (http://theory.sinp.msu.ru/comphep)
there are several new  
improvements incorporated
which make the program significantly more efficient to use for 
event simulation, especially for the case of hadronic collisions.

\subsection{QUARK FLAVOR COMBINATORICS}

A serious computational problem for the hadronic collisions 
is a large number of partonic subprocesses with many contributing
Feynman diagrams in each of them. This takes place, in particular,
 for the reason
of numerous possible combinations of initial and final quarks with 
different flavors, giving different subprocesses. Many additional 
diagrams appear in each of these subprocess due to CKM quark mixing.
In fact many of the subprocesses and many diagrams have very
similar topological and Lorentz structure, so one would like to 
compute all the similar pieces only once.
 
In the paper \cite{Boos:2000ap} the method for a simplification
of quarks combinatorics has been proposed. The method
has been implemented to the CompHEP 4.4 and tested on many
examples showing a significant reduction in a number of 
contributing subprocesses and diagrams.  

The method is based on two approximations:
\bi  
\item{ mixing with the third generation is neglected \\
$ V_{CKM}  \;\Longrightarrow\; \left( \begin{array}{cc}
                                            V & 0 \\
                                            0  & 1
                                       \end{array} \right)\;,\\
   \qquad
   V \;=\;  \left( \begin{array}{cc}
                         \cos\vartheta_c  & \sin\vartheta_c \\
                         -\sin\vartheta_c & \cos\vartheta_c
                                       \end{array} \right)$\\
where $\vartheta_c$ is the Cabbibo angle.}
\item{ masses of the first four quarks are neglected\\
$M_u = M_d = M_s = M_c = 0$}
\ei

One should stress that usually both approximations work very well 
for collider processes, although in the case when masses are needed to be 
taken into account one should use the regular CompHEP Standard Model.

Under the assumptions above the method is based on the two main ideas:
\bi 
\item{rotation of down quarks, thus, transporting the mixing matrix 
elements from the
vertices of subprocess Feynman diagrams to the parton distribution
functions}
\item{the contributing diagrams are splitted into exact gauge invariant 
subclasses \cite{Boos:1999qc} for each of the subprocess 
according to a different topology
of the quark lines involved.}  
\ei

As a result one finds new rules for a convolution with the structure 
functions 
where each  gauge invariant subclass of squared diagrams with certain topology
of the quark lines is convoluted with a corresponding combination
of structure functions. In general there are only three different
possibilities which we refer to as a convolution rules number 1,2 and 3.

{\bf Scattering topology. 1st Rule} 

For this topology there are two quark loops after squaring, and  
each loop includes only one initial quark. 
In such a case the structure
functions are summed over the two flavors for each initial state
and the dependence on Cabbibo angle disappears.
The following formula gives a particular example for this case when one
initial state is down quarks while another initial state is up antiquarks:
\vspace{-3mm}
\begin{eqnarray}
     |{\cal D}_{sc}|^2  &\Longrightarrow &
    \int dx_1 dx_2 ~[f_d(x_1)+f_s(x_1)] \nonumber \\
 && [f_{\bar u}(x_2)+f_{\bar c}(x_2)]~|{\cal M}|^2 \nonumber
\end{eqnarray}
\vspace{-10mm}
  \begin{figure}[hbtp]
\includegraphics[width=30mm,height=30mm,angle=270]{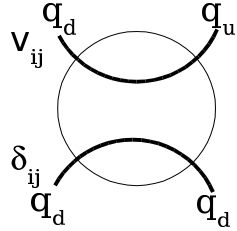}
\vspace{-5mm}
\caption{Scattering Topology (see the Figures for other 
topologies in \cite{Boos:2000ap})}
\label{fig:scat_top}
\end{figure}

\vspace{5mm}
{\bf Annihilation CC topology. 2nd Rule}

For this topology there is only one quark loop in the squared diagram
which includes both initial quarks.
 For the CC case one of the initial state partons is always the up type 
quark
or antiquark while the another initial state parton is the down type quark 
or
antiquark.  
The effective function is the sum of the 
individual quark and antiquark structure function
products times $\cos^2\vartheta_c$ ($\sin^2\vartheta_c$)
in case of the same (different) generation for the quark and antiquark. 
 Some generic example for the convolution in this case is

\vspace{-5mm}
\begin{eqnarray}
 |{\cal D}_{a}^{CC}|^2 &\Longrightarrow &
 \int dx_1 dx_2 ~[f_d(x_1) f_{\bar u}(x_2) \cos^2\vartheta_c \nonumber \\
&&    + f_s(x_1) f_{\bar c}(x_2) \cos^2\vartheta_c  \nonumber\\
&& +f_d(x_1)\,f_{\bar c}(x_2) \sin^2\vartheta_c  \nonumber\\
&&   +f_s(x_1)\,f_{\bar u}(x_2) \sin^2\vartheta_c]~|{\cal M}|^2 \nonumber
\end{eqnarray}

{\bf Annihilation NC topology. 3d Rule}

This topology is similar to the previous one  leading 
to one quark loop in the squared diagram. 
But now both initial quark and antiquark are either up or
down type. Generic example for the convolution with structure
functions in this case is the following (note, that 
there is no dependence on the Cabbibo angle as for the first rule):
\vspace{-2mm}
\begin{eqnarray}
|{\cal D}_{a}^{NC}|^2 &\Longrightarrow & \int dx_1 dx_2~ \nonumber\\
&& [f_d(x_1)\,f_{\bar d}(x_2)
                  +f_s(x_1)\,f_{\bar s}(x_2)]~|{\cal M}|^2 \nonumber
\end{eqnarray}

A special rule is used in the cases when the final quarks and 
antiquarks form a loop in a squared diagram not involving the initial 
quarks.
In these cases each loop of this sort gives a factor of 2
after summation over flavors.
Of course, this rule is valid only under the  assumption that the
fragmentation of the four light quarks and antiquarks leads to
indistinguishable jets. If one includes nontrivial fragmentation
functions, e.g. for a $c$-quark, the above rules have to be modified.

All the above ideas are implemented in CompHEP in the special model
denominated as $SM_{ud}$.
For the processes where only QCD interactions are involved
one can simplify further the light quark flavor combinatorics
since QCD does not distinguish different flavors and electroweak
charges. The corresponding CompHEP model is denominated as $SM_{qQ}$. 

\subsection{EVENT COLOR FLOWS }
As will be described below, the CompHEP has an interface to
PYTHIA program. (The interface to HERWIG is in progress).
For a fragmentation and hadronization PYTHIA needs 
an information on color configurations for the generated partonic 
level events. 
For the color flows PYTHIA requires an indication of colorless 
pairs or, in other words, the definition of {\it color chains}.  
In terms of QCD the sequences of color chains
correspond to the elements of the orthogonal color basis in the
limit of infinite number of colors, $N_c=\infty$.
However CompHEP, as well as other matrix element programs,
calculates exactly the matrix element squared with the number of colors 
$N_c=3$. 

The  procedure described in this section is used in CompHEP in order
to match the $|M|^2$ calculation with $N_c=3$ and the color chains
generation needed for further hadronization.
Note, that originally this algorithm has been implemented in
CompHEP 4.1.10 by A.~Pukhov.

The matrix element is a sum of Feynman diagrams
$$ M \;=\; \sum_d L_d T_d$$
where  $L_d$ is the Lorentz part of the diagram,
and $T_d$ is its color structure .
In the limit $N_c=\infty$ the color structure
of diagrams can be decomposed over the elements of color basis
(color chains) as
$$ T_d \;\Longrightarrow\; T_d^{N_c=\infty} = \sum_i C_d^i t_i$$
where the elements of the basis are orthonormal, $t_i\otimes 
t_j=\delta_{ij}$.
Thus, the squared matrix element can be represented in the limit 
$N_c=\infty$
as a sum of positive contributions
$$  |M|^2_{N_c=\infty} = \sum_i k_i\,,\qquad
    k_i =|\sum_d L_d C_d^i|^2 \,.
$$
As a result, for each event (phase space point) one gets the set of
color chains $t_i$  with weights
$w_i = \frac{k_i}{\sum_j k_j}$.

So the event generation procedure includes two steps. First step is 
the standard generation of an event as a phase space point with $|M|^2$ 
calculated at $N_c=3$ like in the 
standard event weighting procedure. At the second step color chains 
$t_i$ are selected with the weighting factors $k_i$, $N_c=\infty$.

\subsection{LANHEP AND SUSY MODELS}
The LanHEP program \cite{Semenov:2002lh} was developed in the framework of 
the CompHEP project
to implement easily "new physics" models and/or anomalous effective Lagrangian 
terms. LanHEP input is
the Lagrangian written in the compact form close to one used in textbooks.
 In particular, the Lagrangian terms as the LanHEP input can be written 
with summation over indices of the space-time or gauge symmetries and 
using special symbols
for complicated expressions, such as the covariant derivative and  
the strength tensor for gauge fields.
There are 2-component spinors and the superpotential formalism available. 
Correctness of the model can be checked for hermiticity, mass matrix 
diagonalization, and BRST invariance.
The output of the program
is a set of Feynman rules in terms of physical fields and independent 
parameters.
This output can be written in \LaTeX{} format and in the form of CompHEP
model files, which allows one to start calculations of processes in
the new physical model. 

We have used the description of the MSSM physical spectrum and the 
Lagrangian 
for 3 generations of quarks and leptons given in \cite{Rosiek:1990}.
We also add the effective Higgs potential to the tree
level MSSM Lagrangian to incorporate the radiative corrections 
as it is given in the FeynHiggs program \cite{Heinemeyer:1998yj} to
the masses and couplings of Higgs bosons in a gauge invariant way.
Details of the notation for the particles and parameters of the MSSM in 
CompHEP
are given in \cite{Semenov:2002mssm} 
The Feynman rules generated by LanHEP were successfully
compared with the ones listed in \cite{Rosiek:1990}, as well as with many 
results
of other publications. The comparison to GRACE/SUSY \cite{Fujimoto:2002sj}
program was successfully done for many processes.

The SUGRA and GMSB SUSY models are also implemented to the CompHEP by means of
linking with the ISAJET \cite{Paige:2003mg} library. There are also MSSM 
extensions available, such
as R-parity violation model and the model with gravitino and sgoldstino 
fields \cite{Gorbunov:2001sg}. 

Lagrangians for other nonstandard models, e.g. models with 
leptoquarks \cite{Ilyin:1995jv,Blumlein:1996qp}, 
complete two-Higgs-doublet model with conserved or broken CP invariance
\cite{Dubinin:2002nx}, 
models with anomalous top quark interactions 
\cite{Boos:1999ca} are also available for 
CompHEP calculations.

\subsection{COMPHEP-PYTHIA INTERFACE}
PYTHIA has the built-in database of matrix
elements for hard subprocesses which are basically of the $2\to 2$ type.
The CompHEP-PYTHIA interface (the old version of the interface
is described in \cite{Belyaev:2000wn}) allows to use the processes $2\to 
3,4,5,6$ computed by means of CompHEP as the "new" processes for PYTHIA 
and to include in this way ISR/FSR, hadronization, 
and decays like it is done in PYTHIA. 

CompHEP generates unweighted events and writes them to the file
{\it $events\_N.txt$} where N is the number of working session.
The command {\it mixPEV}  (syntax: {\it ./mixPEV $ events{\_N_1}.txt$ 
$ events{\_N_2}.txt$ ...}) mixes randomly several event files corresponding to 
some subprocesses in one event flow according to their
relative weights, $\sigma_i/\Sigma_j\sigma_j$, where $\sigma_i$ is the
cross section of the $i$-th subprocess, and writes events to the file
 {\it Mixed.PEV} . This file is ready for input to 
PYTHIA 6.2 and contains the data necessary for Les Houches Accord I
\cite{Boos:2001cv}.

We provide the interface library  
{\it libinterface62.a}  and an example of 
the  {\it main.f}  program which is very easy to be incorporated
to any standard PYTHIA main.f program. Using the
standard Unix  "make"  command
a user links main.o, libinterface62.a and pythia62.o 
to the executable file {\it generator.exe} . 
The generator takes events from the file Mixed.PEV.
The number of requested events to be generated after
PYTHIA is defined by user in the file {\it INPARM.DAT}

There are already many examples of the CompHEP and CompHEP-PYTHIA
interface use for the real analysis at the Tevatron, LHC and LC.

Of course, as pointed out in the introduction, one has to be careful 
combining new higher order processes calculated with CompHEP-PYTHIA 
interface, and the standard PYTHIA processes.
Double counting of contributions should be carefully analysed,
not only when ISR/FSR are switched on, but in many other cases.

\subsection{COLLIDING BEAMS IN COMPHEP}
 New option is implemented in CompHEP 4.4 - 
 a possibility to enter a {\it beam} as
the initial state. In CompHEP the beam is a set of partons. 
 This is a natural terminology for hadronic collisions
(proton or antiproton beams). In CompHEP 4.4 this option allows 
a user to enter also an electron, photon or quark itself
as a {\it beam}.  In this way one can set, for example, the photon as a 
"parton" in electron (effective photon approximation), 
quarks and gluons as "partons" in a photon
(resolved photons), W-boson as a "parton" in electron or quark
(effective W-approximation) etc.   
This option is specially useful for lepton and lepton-hadron colliders.
Moreover, one can introduce a beam with arbitrary content for any private
purposes.

 The parton distribution functions are assigned now to the beam by a user
before the process is entered. All predefined beams are collected into a 
special
table which can be modified by a user if necessary. When the user sets
a process (s)he indicates a particle or a beam (from the table) as an
initial state. The corresponding table can be displayed on the screen by
pressing the F3 key.

In the CompHEP 4.4 the latest versions of the CTEQ parton distribution
functions \cite{Pumplin:2002vw} are implemented. However, some previous
sets are also available since for an analysis where they could be needed.

\subsection{SYMBOLIC BATCH MODE}
  The main goal of the symbolic batch script is to launch the symbolic
calculation of squared diagrams  using a command line interface
(the non-GUI mode). This  allows the reuse
of the script, avoiding typing mistakes, and also helps to send large
symbolic calculation tasks to the computer farms.

  The script imitates the typing input data in CompHEP menu (initial
states, energies, final particles etc). This option was described in
\cite{Pukhov:2002} - so-called {\it blind mode} in CompHEP 4.1.10.

All  the data, necessary for further symbolic calculations, are taken from
a file with a default name {\it process.dat}. Detailed instructions and
explanations for the parameters are contained in this file.

After filling in the file process.dat a user launches the script. The
results of the script execution will be the  binary file {\it
s\_comphep.exe} and the file {\it symb\_batch.log} where all details on
the binary creation are stored.

One can set another name for the data file or the "results" directory
by the -n and -d options respectively. If the CompHEP directory "results"
is not empty before the script is launched, it is renamed to
{\it results\_old\_0} and the script creates a new "results" directory.
The symbolic batch script has some extra options:

\begin{itemize}

\item[] {\bf -recovery}
if the s\_comphep program crashes during calculations of squared diagrams,
one can launch the script with the recovery option, and the s\_comphep resumes
computations from the last calculated diagram;

\item[] {\bf -relink}
if the user has changed the userFun.c file and wants to relink the 
n\_comphep
program with a new file, (s)he can launch the script with the relink
option. All details of the relinking are stored in the symb\_batch.log
file;

\item[] {\bf -show diagrams}
this option is applied if the user wants to exclude by hand some diagrams
from the calculation.
The script launches s\_comphep in GUI mode in
the Feynman diagrams menu. After reviewing the diagrams the user has to finish
the GUI session and the script will go on.
\end{itemize}

\subsection{NUMERICAL BATCH MODE}
  The CompHEP was originally created as a program with an interactive
GUI interface. However in practice for many cases 
a command line interface (non-GUI) is very useful
and it was implemented in the CompHEP 4.4 numerical batch mode.
The interface
allows to perform large time scale calculations using,
if necessary, the computer farms (PBS and LSF systems), to compute 
processes with a large number of subprocesses, in particular, 
in parallel if computer resources are enough.
The interface is very useful for repeating numerical calculations
by non-expert users without making trivial mistakes.  

In this mode numerical computations start
after finishing the symbolic computations and creation of  
the executable file n\_comphep.exe in the directory "results".
At the first step a user has to create the numerical batch data file
batch.dat by launching the command $  ./num\_batch.pl $  from the
user working directory, where CompHEP has been started. The created
file batch.dat contains the default parameters 
(model parameters, cuts, kinematics, etc.) listed in the file
session.dat in the directory "results".
If necessary the user can change any of these parameters by editing
the file batch.dat or using GUI interface to modify initial session.dat file.
In the latter case one can start the command 
$  ./num\_batch.pl$ with the option $ --add\_ses2bat $ to include parameters 
from the new session.dat file to the batch.dat.  

The command (script) $  ./num\_batch.pl$ has many other useful options,
for example:
\begin{itemize}
\vspace{-0.2cm} \item[] {\bf --help} - to list the complete set of options with an explanation 
\vspace{-0.2cm} \item[] {\bf -d dir\_name} - to change the default directory results 
to the directory dir\_name
\vspace{-0.2cm} \item[] {\bf --add\_ses2bat [file]}  - to add parameters from existing 
file session.dat in the directory results or from [file], if indicated, to the batch.dat file 
\vspace{-0.2cm} \item[] {\bf -run [vegas,max,evnt]} - to start numerical calculations 
including all steps by default or only the steps as indicated by the keys
[vegas,max,evnt], where [vegas] means the cross section calculation,
[max] - search for maxima in each of the phase space sub-cubes, [evnt] - unweighted 
event generation
\vspace{-0.2cm} \item[] {\bf -run [cleanstat,cleangrid,clean]} - to clear statistics or grid 
or both 
\vspace{-0.2cm} \item[] {\bf -proc n1,n2-n3,n4...}  - to run only the subprocesses with
corresponding numbers
\vspace{-0.2cm} \item[] {\bf -pbs [``pbs prefix'']}  - to start calculations in 
parallel with PBS batch system. The default PBS prefix is: `` qsub -I ''  
\vspace{-0.2cm} \item[] {\bf -lsf [``lsf prefix'']} -  to start calculations in 
parallel with LSF batch system. The default LSF prefix is: ``bsub -I ''
\end{itemize}

For example, the command \\
$      ./num\_batch.pl  \ -run \ vegas \ -pbs $ \\
 creates temporary subdirectories for each of the subprocess
and run the PBS batch job for each of subprocess with PBS command:
\mbox{qsub -I run.sh}, where the script run.sh is created automatically
and starts the n\_comphep.exe.
These temporary subdirectories will be automatically removed and 
the results for all subprocess will be saved in one directory. 
   
A  user can also calculate subprocesses in parallel without the PBS (LSF) batch
system by starting the same command with empty PBS (LSF) prefix:\\
$      \mbox{./num\_batch.pl  \ -run \ vegas \ -lsf \ ' \ '} $ \\

\subsection{FUTURE PLANS}
There are several directions  for the future development of
CompHEP.  One important direction is a further development of
distributed Monte Carlo calculation and event generation
on computer clusters ~\cite{Kryukov:2003}. 
 Another important direction is an implementation of the 
 FORM computer algebra  program \cite{Vermaseren} for symbolic 
 calculations of matrix elements (some
preliminary realization was described in \cite{Bunichev:xd}).
This option will allow to introduce new complicated structures in the
vertices (e.g. form-factors), implement new algorithms to increase
the efficiency of symbolic calculations, perform calculations
in theories with extra dimensions, use the dimensional regularization,
perform polarized calculations by introducing the corresponding 
density matrices for external lines of squared diagrams.
We plan to create the Les Houches accord 1 based interface to HERWIG
and the Les Houches accord 2 based interface to PDFs
as well as the SUSY  Les Houches accord based interface 
\cite{Skands:2003cj} to various codes calculating the SUSY 
mass spectrum. 

The long term CompHEP collaboration projects which
are under discussion include symbolic
and numerical amplitude calculations with extension to the
1-loop case using various methods, incorporation the gauge
invariant classes of diagrams, etc.   
\section{THE COMPHEP BASED GENERATOR SINGLETOP}
The CompHEP has been used to prepare a special event generator
SingleTop to simulate the electroweak single top quark 
production with its subsequent decays at the Tevatron and LHC.
Single top is expected to be discovered at the Tevatron Run II
and will a very interesting subject of detail studies at
the LHC (see the reviews \cite{Beneke:2000hk,lhc_lc}).

There are three main processes of singe top production at hadron
colliders characterizing by the virtuality $Q^2_W$
of the participating at the process $W$-boson: t-channel,
s-channel and associated $tW$ mechanisms respectively.

\begin{figure}[hbtp]
\includegraphics[width=75mm,height=20mm]{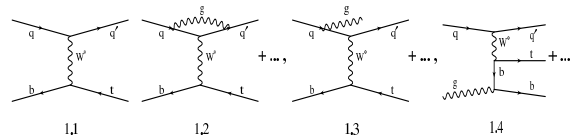}
\vspace{-15mm}
\caption{LO order and representative loop and tree NLO
diagrams to the t-channel single top production }
\label{SinT}
\end{figure}

The generator SingleTop includes all the three processes and 
provides Monte-Carlo unweighted events at the NLO QCD level.
We discuss shortly here only the main process with the largest 
rate, the t-channel production.
The representative NLO diagrams are shown in  Fig.\ref{SinT} 
The top decay is not shown, however it is included with all the 
spin correlations.  

We compute by means of the CompHEP the LO order process $2\rightarrow 2$
with the b-quark in the initial state and top spin correlated
$1\rightarrow 3$ subsequent decay, put it into PYTHIA using the interface
and switch on ISR/FSR. Then with CompHEP we compute   
the NLO tree level corrections - $2\rightarrow 3$ processes with 
additional b- and light quarks or gluons in the final state 
including also the top decay with spin correlations.
We split the phase space region in "soft" and "hard" parts on $p_t$ 
of those additional b and light jets being from PYTHIA radiation
in the "soft" and from the CompHEP matrix element calculation
in the "hard" regions. The soft part is normalized in such a way that 
all parts being taken together give
known from  calculations the NLO cross section \cite{Stelzer:1997ns,Harris:2002md}.
The splitting parameters are turned such that all the distribution
become smooth after the normalization. The performed
cross checks show that the computed NLO distributions \cite{Harris:2002md} are 
correctly reproduced. Therefore, prepared in that way generator
does not have a double counting, gives the NLO rate and distributions,
and include all the spin correlations. 

The first release of the generator \cite{cmsnote} did not include
the hard radiation of the light jets, while the latest version
\cite{SingleTop} currently used in the analysis by the Fermilab DO 
and the LHC CMS collaborations includes all the mentioned properties.

There are several recent examples of the CompHEP use for various
simulations, namely, 
the generator for MSSM Higgs bosons in the intense coupling regime
at the LHC \cite{Boos:2003jt}, the generator for sfermion pair production
with their subsequent decays to polarized fermions \cite{Boos:2003vf}
and the generator for measurements of the $HZ\gamma$ coupling
of \cite{Dubinin:2003kb} at Linear Colliders.

\section{CONCLUSIONS}
The CompHEP package 
is a powerful tool for a simulation of different
physical processes at hadron and lepton colliders.
The most important CompHEP 4.4 improvements include the build-in
MSSM, simplification of quark flavor combinatorics, generation
of unweighted events with color chains defined for the following
showering and hadronization procedure, 
Les Houches accord based interface with PYTHIA, symbolic and numerical
batch modes, the latest PDF sets implementation and
a new treatment of the initial beams.

The CompHEP is basically the LO program.  However it allows
to include partly the NLO corrections. NLO tree level $2\rightarrow N+1$
corrections to the $2\rightarrow N$ process can be computed.
One can include the NLO structure functions and loop relations between 
the parameters like in the MSSM,
K-factors when available, and loop contributions from the existing 
publications as form-factors.
However a correct matching to complete NLO is obviously a 
nontrivial problem
which has to be carefully considered in each particular physical case.
The CompHEP based generator SingleTop is an example of a reasonable
matching of the different-order contributions.

\section{ACKNOWLEDGMENTS}
The authors thank the Russian Ministry of Industry, Science
and Technology for the support of the CompHEP project.
We acknowledge the support of the RFBR grant 04-02-17448
and the grant of the program "University of Russia".
We are grateful to our colleagues for using the CompHEP,
for reporting us on problems and bugs.

 We specially thank the
Organizing Committee of the ACAT2003 Conference for 
 a support.
 


\begin{thebibliography}{9}
\bibitem{Pukhov:1999gg}
A.~Pukhov, E.~Boos, M.~Dubinin, V.~Edneral, V.~Ilyin, D.~Kovalenko, 
A.~Kryukov, 
V.~Savrin, S.Shichanin, A.~Semenov,
arXiv:hep-ph/9908288. \\
Web cite: http://theory.sinp.msu.ru/comphep

\bibitem{Fujimoto:2002sj}
 J.~Fujimoto, T.~Ishikawa, M.~Jimbo, T.~Kaneko, K.~Kato, 
 S.~Kawabata, K.~Kon , M.~Kuroda, Y.~Kurihara, Y.~Shimizu, H.~Tanaka 
Comput.\ Phys.\ Commun.\  {\bf 153}, 106 (2003)
[arXiv:hep-ph/0208036];

\bibitem{Stelzer:1994ta}
T.~Stelzer and W.~F.~Long,
Comput.\ Phys.\ Commun.\  {\bf 81}, 357 (1994)
[arXiv:hep-ph/9401258];
F.~Maltoni and T.~Stelzer,
JHEP {\bf 0302}, 027 (2003)
[arXiv:hep-ph/0208156].

\bibitem{Mangano:2002ea}
M.~L.~Mangano, M.~Moretti, F.~Piccinini, R.~Pittau and A.~D.~Polosa,
JHEP {\bf 0307}, 001 (2003)
[arXiv:hep-ph/0206293].

\bibitem{Moretti:2001zz}
M.~Moretti, T.~Ohl and J.~Reuter,
arXiv:hep-ph/0102195.

\bibitem{Kilian:2002cg}
W.~Kilian,
{\it Prepared for 31st International Conference on High Energy 
Physics (ICHEP 2002), Amsterdam, The Netherlands, 24-31 Jul 2002}

\bibitem{Krauss:2001iv}
F.~Krauss, R.~Kuhn and G.~Soff,
JHEP {\bf 0202}, 044 (2002)
[arXiv:hep-ph/0109036].

\bibitem{Kanaki:2000ms}
A.~Kanaki and C.~G.~Papadopoulos,
arXiv:hep-ph/0012004.

\bibitem{Campbell:2003hd}
J.~Campbell, R.~K.~Ellis and D.~Rainwater,
Phys.\ Rev.\ D {\bf 68}, 094021 (2003)
[arXiv:hep-ph/0308195].

\bibitem{Beenakker:2002nc}
W.~Beenakker, S.~Dittmaier, M.~Kramer, B.~Plumper, M.~Spira and P.~M.~Zerwas,
Nucl.\ Phys.\ B {\bf 653}, 151 (2003)
[arXiv:hep-ph/0211352].

\bibitem{Dawson:2003zu}
S.~Dawson, C.~Jackson, L.~H.~Orr, L.~Reina and D.~Wackeroth,
Phys.\ Rev.\ D {\bf 68}, 034022 (2003)
[arXiv:hep-ph/0305087].

\bibitem{Belanger:2003nm}
G. Belanger, F. Boudjema, J. Fujimoto, T. Ishikawa, T. Kaneko, K. Kato, 
Y. Shimizu, Y. Yasui.
Phys.\ Lett.\ B {\bf 571}, 163 (2003)
[arXiv:hep-ph/0307029].

\bibitem{Sjostrand:2003wg}
T.~Sjostrand, L.~Lonnblad, S.~Mrenna and P.~Skands,
arXiv:hep-ph/0308153.

\bibitem{Corcella:2000bw}
G. Corcella, I.G. Knowles, G. Marchesini, S. Moretti, K. Odagiri, 
P. Richardson, M.H. Seymour, B.R. Webber. 
JHEP {\bf 0101}, 010 (2001)
[arXiv:hep-ph/0011363].

\bibitem{Boos:2001cv}
E.~Boos, M.~Dobbs, W.~Giele, I.~Hinchliffe, J.~Huston, V.~Ilyin, 
J.~Kanzaki, K.~Kato, Y.~Kurihara, L.~Lonnblad, M.~Mangano, 
S.~Mrenna, F.~Paige, E.~Richter-Was, M.~Seymour, 
T.~Sjostrand, B.~Webber, D.~Zeppenfeld
arXiv:hep-ph/0109068.

\bibitem{Catani:2001cc}
S.~Catani, F.~Krauss, R.~Kuhn and B.~R.~Webber,
JHEP {\bf 0111}, 063 (2001)
[arXiv:hep-ph/0109231].

\bibitem{Frixione:2002ik}
S.~Frixione and B.~R.~Webber,
JHEP {\bf 0206}, 029 (2002)
[arXiv:hep-ph/0204244].

\bibitem{Frixione:2004wy}
S.~Frixione and B.~R.~Webber,
arXiv:hep-ph/0402116.

\bibitem{Mrenna:2003if}
S.~Mrenna and P.~Richardson,
arXiv:hep-ph/0312274.

\bibitem{Mangano:2003ps}
M.~L.~Mangano,
eConf {\bf C030614}, 015 (2003)
[arXiv:hep-ph/0312117].

\bibitem{Baikov:1997zr}
P.~Baikov, E.~Boos, M.~Dubinin, V.~Edneral, V.~Ilyin,
 D.~Kovalenko, A.~Kryukov, A.~Pukhov, V.~Savrin, 
 A.~Semenov, S.~Shichanin.
arXiv:hep-ph/9701412.

\bibitem{Belyaev:2000wn}
A.~Belyaev, E.~Boos, A.~Vologdin, M.~Dubinin, V.~Ilyin,
A.~Kryukov, A.~Pukhov, A.~Skachkova, V.~Savrin, 
 A.~Sherstnev, S.~Shichanin. Oct 2000. 4pp.
arXiv:hep-ph/0101232.

\bibitem{Boos:2000ap}
E.~E.~Boos, V.~A.~Ilyin and A.~N.~Skachkova,
JHEP {\bf 0005}, 052 (2000)
[arXiv:hep-ph/0004194].

\bibitem{Boos:1999qc}
E.~Boos and T.~Ohl,
Phys.\ Rev.\ Lett.\  {\bf 83}, 480 (1999)
[arXiv:hep-ph/9903357].

\bibitem{Semenov:2002lh}
 A.~Semenov, LAPTH-926/02, Annecy, 2002; arXiv:hep-ph/0208011.

\bibitem{Rosiek:1990} J.Rosiek, Phys. Rev. D41, 3464 (1990),
{\it errata} arXiv:hep-ph/9511250.

\bibitem{Heinemeyer:1998yj}
S.~Heinemeyer, W.~Hollik and G.~Weiglein,
Comput.\ Phys.\ Commun.\  {\bf 124}, 76 (2000)
[arXiv:hep-ph/9812320].

\bibitem{Semenov:2002mssm}
 A.~Semenov, Nucl.Inst.Meth. A502 (2003) 558; arXiv:hep-ph/0205020.

\bibitem{Paige:2003mg}
F.~E.~Paige, S.~D.~Protopescu, H.~Baer and X.~Tata,
arXiv:hep-ph/0312045.

\bibitem{Gorbunov:2001sg}
D.~Gorbunov, A.~Semenov, LAPTH-885/01, arXiv:hep-ph/01111291.

\bibitem{Ilyin:1995jv}
V.~A.~Ilyin, A.~E.~Pukhov, V.~I.~Savrin, A.~V.~Semenov and W.~B.~von Schlippe,
Phys.\ Lett.\ B {\bf 351}, 504 (1995)
[Erratum-ibid.\ B {\bf 352}, 500 (1995)]
[arXiv:hep-ph/9503401].


\bibitem{Blumlein:1996qp}
J.~Blumlein, E.~Boos and A.~Kryukov,
Z.\ Phys.\ C {\bf 76}, 137 (1997)
[arXiv:hep-ph/9610408].

\bibitem{Dubinin:2002nx}
M.~N.~Dubinin and A.~V.~Semenov,
Eur.\ Phys.\ J.\ C {\bf 28}, 223 (2003)
[arXiv:hep-ph/0206205].

\bibitem{Boos:1999ca}
E.~Boos, M.~Dubinin, M.~Sachwitz and H.~J.~Schreiber,
Eur.\ Phys.\ J.\ C {\bf 16}, 269 (2000)
[arXiv:hep-ph/0001048].

\bibitem{Pumplin:2002vw}
J.~Pumplin, D.~R.~Stump, J.~Huston, H.~L.~Lai, P.~Nadolsky and W.~K.~Tung,
JHEP {\bf 0207}, 012 (2002)
[arXiv:hep-ph/0201195].

\bibitem{Pukhov:2002}
A.~Pukhov
Nucl.\ Instrum.\ Meth.\ A {\bf 502}, 327 (2003).

\bibitem{Kryukov:2003}
A.~Kryukov, L.~Shamardin. In this proceedings.

\bibitem{Vermaseren}
J.A.M.~Vermaseren,
KEK-PREPRINT-92-1, Mar 1992. 20pp.;
arXiv:math-ph/0010025;
J.~A.~M.~Vermaseren,
Nucl.\ Phys.\ Proc.\ Suppl.\  {\bf 116}, 343 (2003)
[arXiv:hep-ph/0211297].

\bibitem{Bunichev:xd}
V.~Bunichev, A.~Kryukov and A.~Vologdin,
Nucl.\ Instrum.\ Meth.\ A {\bf 502}, 564 (2003).

\bibitem{Skands:2003cj}
P.~Skands {\it et al.},
arXiv:hep-ph/0311123.

\bibitem{Beneke:2000hk}
M.~Beneke {\it et al.},
arXiv:hep-ph/0003033.

\bibitem{lhc_lc}
LHC/LC Study Group Working Document. In preparation

\bibitem{Stelzer:1997ns}
T.~Stelzer, Z.~Sullivan and S.~Willenbrock,
Phys.\ Rev.\ D {\bf 56}, 5919 (1997)

\bibitem{Harris:2002md}
B.~W.~Harris, E.~Laenen, L.~Phaf, Z.~Sullivan and S.~Weinzierl,
Phys.\ Rev.\ D {\bf 66}, 054024 (2002)

\bibitem{cmsnote}
E.~Boos, L.~Dudko and V.~Savrin, CMS Note 2000/065

\bibitem{SingleTop}
E.~Boos, L.~Dudko,V.~Savrin, and A.~Sherstnev, in preparation 

\bibitem{Boos:2003jt}
E.~Boos, A.~Djouadi and A.~Nikitenko,
Phys.\ Lett.\ B {\bf 578}, 384 (2004)
[arXiv:hep-ph/0307079].

\bibitem{Boos:2003vf}
E.~Boos, H.~U.~Martyn, G.~Moortgat-Pick, M.~Sachwitz, A.~Sherstnev and P.~M.~Zerwas,
Eur.\ Phys.\ J.\ C {\bf 30}, 395 (2003)
[arXiv:hep-ph/0303110].

\bibitem{Dubinin:2003kb}
M.~Dubinin, H.~J.~Schreiber and A.~Vologdin,
Eur.\ Phys.\ J.\ C {\bf 30}, 337 (2003)
[arXiv:hep-ph/0302250].

\end{thebibliography}
\end{document}